## Title

- Solving the cooling flow problem with combined jet-wind AGN feedback

## Authors

Aoyun He,[1,2] Feng Yuan,[3]* Suoqing Ji,[3,4]* Minhang Guo,[1,5,2] Yuan Li,[6] Haiguang Xu,[7] Ming Sun,[8] Haojie Xia,[1,2] Yuanyuan Zhao[7]

## Affiliations

[1]Shanghai Astronomical Observatory, Chinese Academy of Sciences, 80 Nandan Road, Shanghai 200030, China.
[2]University of Chinese Academy of Sciences, No. 19A Yuquan Road, Beijing 100049, China.
[3]Center for Astronomy and Astrophysics and Department of Physics, Fudan University, Shanghai 200438, China.
[4]Key Laboratory of Nuclear Physics and Ion-Beam Application (MOE), Fudan University, Shanghai 200433, China.
[5]ShanghaiTech University, 393 Middle Huaxia Road, Shanghai 201210, China.
[6]Department of Astronomy, University of Massachusetts, 710 North Pleasant Street, Amherst, MA 01003-9305, USA.
[7]State Key Laboratory of Dark Matter Physics, School of Physics and Astronomy, Shanghai Jiao Tong University, 800 Dongchuan Road, Shanghai 200240, China.
[8]Department of Physics & Astronomy, University of Alabama in Huntsville, Huntsville, AL 35899, USA.

***Corresponding author. Email: fyuan@fudan.edu.cn, sqji@fudan.edu.cn**

## Abstract

Active galactic nucleus (AGN) feedback is widely viewed as the most promising solution to the long-standing cooling flow problem in galaxy clusters, yet previous models prescribe jet properties inconsistent with accretion physics. We perform an idealized hydrodynamic simulation of a galaxy cluster with no merger history and a relaxed state, with its other properties similar to the Perseus cluster using the MACER framework, incorporating both jets and winds whose properties are constrained by general relativistic magnetohydrodynamic simulations of black hole accretion and observations. The combined feedback reproduces key observables, including cold gas mass, star formation rate, thermodynamic radial profiles, and black hole growth, while jet-only or wind-only models fail. The success arises from turbulence driven by jet–wind shear that enhances kinetic-to-thermal energy conversion, boosting heating efficiency by factors of three and six relative to wind-only and jet-only cases, respectively.

## Teaser

Combined wind and jet feedback self-consistently solves the cooling flow problem in galaxy clusters.

## MAIN TEXT

**The manuscript should be a maximum of 15,000 words.**

### Introduction

A large fraction of galaxy clusters exhibit a cool-core (CC) structure, where the central gas is characterized by high density and low entropy, with cooling timescales on the order of ~1 Gyr (*1*). Under such conditions, a strong cooling inflow at rates of $10^2 - 10^3 M_\odot$ yr $^{-1}$ is expected to develop, accompanied by the accumulation of large amounts of cold gas ($\sim 10^{10} - 10^{11}\ M_\odot$) in cluster cores and high star formation rates (SFRs) of hundreds or even thousands of solar masses per year in the central brightest cluster galaxies (BCGs) (*2*). However, multi-band observations have found far less cold gas ($\sim 10^8 - 10^{10} M_\odot$) and smaller star formation rate ($0.1 - 10s\ M_\odot$ yr $^{-1}$) than predicted (*3*). This discrepancy is referred to as the "cooling flow problem."

To reconcile this discrepancy, many mechanisms have been proposed, such as stellar and supernova feedback (*4*), cosmic rays (*5*), magnetic fields (*6*), and thermal conduction (*7*). Among them, active galactic nuclei (AGNs) feedback has emerged as the dominant solution (*8-20*), supported by strong observational evidence linking AGN activity to suppressed cooling, X-ray cavities, and multiphase gas in low-entropy cores (*21-23*). However, the AGN feedback models adopted in existing simulations commonly suffer from important physical drawbacks.

First, due to limited resolution, almost all simulations cannot resolve the outer boundary of the accretion flow, commonly referred to as the Bondi radius, where the gravitational energy of the gas is comparable to its thermal energy; thus, the black hole accretion rate cannot be reliably calculated (*24,25*). Second, with jets (or winds) being invoked to be the main component of AGN feedback, their parameters are usually treated as free, including the opening angle, velocity, and mass flux (*9,12,15*), and the adopted parameter values are often inconsistent with the theories of black hole accretion and jet formation. Even when the jet parameters are treated as free, some studies adopt jet precession or reorientation to enhance the interaction with the Intracluster Medium (ICM) (*11,14,18*). Third, although theoretical and observational works have established that wind are generally expected to be present when jets are launched (*26-30*), most studies to date have not simultaneously and self-consistently incorporated both jet and wind. Finally, many existing idealized cooling flow simulations (e.g., *13,14*) focus primarily on addressing the cooling flow problem itself, without examining whether the proposed solutions are also consistent with other observational constraints, such as the growth of the black hole, which is more commonly tracked in cosmological simulations (e.g., *31*).

### Results

#### Simulations of the evolution of Perseus-like cluster with MACER

In this paper, we perform idealized high-resolution three-dimensional hydrodynamic simulations of the evolution of a galaxy cluster with no merger history, no cosmological inflow, and a relaxed state, with its other properties similar to the Perseus cluster. The simulations are based on the MACER framework, a model we have developed to investigate the evolution of a single galaxy, focusing on the role of AGN feedback (*32,33*; see Methods for details of the simulations). In addition to AGN feedback, other important physical processes such as star formation, stellar feedback, radiative heating and cooling of the ICM by the AGN are also included in MACER. A schematic diagram of the simulation is shown in Figure 1.

Compared with previous idealized cooling-flow simulations and cosmological simulations, the MACER framework has several distinctive features. A detailed comparison is presented in the Methods section, here we briefly summarize the most salient ones. First, the inner boundary of our simulation domain is smaller than the outer boundary of the black hole accretion flow, whose value is shown in the bottom panel of Figure 2. In this case, by combining the mass flux calculated at the inner boundary with black hole accretion theory, we can obtain a reliable estimate of the mass accretion rate at the black hole horizon. This is crucial for studying AGN feedback, since the accretion rate determines the power of each component of AGN outputs, including the jet.

Second, state-of-the-art accretion physics is adopted in MACER. The black hole accretion is divided into hot and cold modes, depending on whether the accretion rate is below or above 2% of the Eddington rate. This boundary value is based on observations of the state transition in black hole X-ray binaries (*28*). In each mode, we consider all AGN outputs, namely radiation, jets, and winds. Specifically, in the hot accretion mode, jet and wind are incorporated simultaneously. Moreover, the parameter values of these outputs as functions of accretion rate, such as the opening angle, velocity, mass flux, and their angular distributions, are not treated as free parameters. Instead, they are directly adopted from, or extrapolated based on, theoretical and observational studies on black hole accretion (*28,29;* See Methods for details). This contrasts with other simulation works and ensures that the winds and jets adopted in our model are physically motivated. As we will show later, the coexistence of wind and jet plays an important role in our feedback model, as the coupling between them produces strong turbulence whose dissipation efficiently converts AGN kinetic energy into the thermal energy of the ICM in the galaxy cluster.

**Simulation results and comparison with observations**

To assess the importance of including both wind and jet, we conduct three simulations that consider different AGN components in the hot mode: `JetWind`, `JetOnly`, and `WindOnly`. The `JetWind` simulation includes both jet and wind, while `JetOnly` and `WindOnly` disable the wind and jet, respectively. When comparing `JetOnly` and `WindOnly` with `JetWind`, the total outflow power at the same mass accretion rate is kept identical. For instance, in the `WindOnly` simulation, the jet power is redistributed into the wind channel, ensuring that the removal of one component does not reduce the total feedback power.

During the first 200 Myr, there is almost no cold gas or star formation, and correspondingly the AGN is very weak. This arises because only hot gas is included in the initial conditions of our simulations. After ~200 Myr, a cooling flow develops, leading to the formation of cold gas through radiative cooling and a corresponding rise in the SFR. The infall of cold gas onto the black hole triggers strong AGN activity and feedback in all three models.

The feedback effects differ among the three models, as shown in Figure 2. The `JetOnly` and `WindOnly` simulations are less effective than `JetWind` in suppressing cold gas and star formation. The `JetWind` model yields the lowest cold gas mass and SFR, followed by `WindOnly` and then `JetOnly`. The cold gas mass and SFR in both the `JetWind` and `WindOnly` models are consistent with observations, whereas the JetOnly model overpredicts them, producing too much cold gas and too high an SFR during most of the evolution. However, if we consider the statistical results of many Perseus-like cooling-flow sources, the WindOnly model would fall towards the extreme cases in that sample (*36*).

While AGNs in most galaxy cluster centers are in the hot mode, observations show that a small fraction appear to be in the cold mode (i.e., quasars). The evolution of AGN luminosity shown in Figure 2 is consistent with this observational result.

Although the SFR and cold gas mass in the WindOnly model are consistent with observations, Figure 2 shows that the black hole mass in this model (and in the JetOnly model) grows to nearly (and above) $10^{10} M_\odot$ over the 1.5 Gyr evolution. Such values substantially exceed the observed black hole mass for the Perseus cluster, which ranges from $3.4 \times 10^8 M_\odot$ (*37*) to $1.1 - 1.2 \times 10^9 M_\odot$ (*38*). The excessive black hole growth in the WindOnly and JetOnly models arises from inefficient feedback, allowing too much cold gas to form and accrete. In contrast, in the JetWind simulation, the black hole mass remains at a reasonable value throughout the evolution, indicating a more effective feedback regulation. We will discuss the physical reasons for this later.

In addition to cold gas mass, SFR, and black hole mass, another important observational constraint is the radial profile of the gas thermodynamic quantities. Figure 3 shows the radial profiles of density-weighted entropy by the three models, along with observational data for comparison (*39,40*). The radial profiles of density and temperature with observational data for three models are plotted in Figure S2. The curves are color-coded to represent simulation time. For the JetOnly model, within ~100 kpc, the entropy and temperature are clearly below the observed values, while the density is higher than observed during most of the simulation, indicating that the jet alone fails to adequately heat the ICM in the cluster core. Between 100 and 200 kpc, at late stages of the simulation, moderate increases in both entropy and temperature are observed, suggesting that feedback can affect the cluster outskirts (*41*).

For the WindOnly model, during most of the simulation time, the predicted entropy is substantially higher than observed (first row), due to overly high temperatures and low densities (see Fig. S2). Unlike the JetOnly case, regions beyond 100 kpc remain almost unaffected, indicating that wind feedback primarily influences the central ICM regions. This occurs because the wind's opening angle is much larger than that of the jet and its power is smaller, so the wind's energy is deposited closer to the AGN.

Figure 2 shows the time evolution of several key quantities in the three models, including the star formation rate (SFR), cold gas mass, AGN bolometric luminosity, black hole mass ($M_{BH}$) and its corresponding Eddington luminosities ($L_{\rm Edd}$) and outer boundary of the accretion flow ($R_{\rm out}$). In the multi-phase ICM environment, each phase has its corresponding $R_{\rm out}$. The outer boundary shown in the figure is the mass flux-weighted value of different phases. The black dashed line represents the observed SFR in NGC 1275 (the BCG in Perseus) measured by (*34*), with the 68% plausible interval indicated by the grey shaded region. Additional constraints on the cold gas mass from (*35*) are also shown. These observational measurements are "snapshot" estimates for this source and should be regarded as upper limits, since the statistical averages for cooling-flow systems are typically lower (*34*).

In contrast, the `JetWind` model reproduces the correct radial profiles of entropy, density, and temperature across the entire simulation domain, consistent with observations of cool-core clusters (*39,40*). Combined with the results in Figure 2, we conclude that the `JetWind` model's predictions are consistent with all key observations, successfully resolving the cooling problem in galaxy clusters.

We further examine two additional diagnostics. The first one is the ratio of the cooling time to the free-fall time. Many analytical and simulation studies have shown that when the minimum value of $t_{\rm cool}/t_{\rm ff}$ falls below a certain threshold $t_{\rm cool}/t_{\rm ff} < 10$ (*10,42,43*), thermal instability is triggered, leading to the formation of large amounts of cold gas. Figure 4 shows the radial profiles of $t_{\rm cool}/t_{\rm ff}$, weighted by density, predicted by the three models. The grey shaded regions represent observed values from the core regions (within 100 kpc) of cool-core galaxy clusters in the ACCEPT database (*44*). Only the `JetWind` model agrees with the observations, while the `JetOnly` and `WindOnly` models predict ratios that lie below and above the observed range, respectively.

We also analyze the turbulence properties predicted by the three models. In the `JetWind` model, the velocity power spectrum within 30 kpc closely reproduces the slope and shape of the short-wavelength portion of the Hα velocity structure function observed by (*45*), while the average velocity dispersion of 150–200 km/s quantitatively matches the X-ray observations of (*46*). The `JetOnly` and `WindOnly` models fail to produce results consistent with observations.

**Interpretation: why only the jet-wind model can solve the cooling flow problem**

Whether the cooling flow problem can be resolved depends on the competition between heating and cooling in the ICM. To understand why only the `JetWind` model succeeds, we first examine the energetics. Although the AGN power in the `JetWind` model is the lowest among the three models, its cooling is also the weakest, leading to the smallest amount of cold gas (Figure 2). We have calculated the total AGN energy and the cooling energy of the ICM released during the whole evolution for the three models. The values are $1.9 \times 10^{62}$ erg and $1.8 \times 10^{62}$ erg for `JetWind`, $4.2 \times 10^{62}$ erg and $6.6 \times 10^{62}$ erg for `WindOnly`, and $1.4 \times 10^{63}$ erg and $2.9 \times 10^{63}$ erg for `JetOnly`, respectively. So among the three models, only in `JetWind` does the time-integrated AGN power exceed the cumulative ICM cooling loss.

This outcome arises from the self-regulated nature of the system. A higher ratio of total cooling to heating energy integrated over some period leads to more cold gas, which in turn enhances accretion and AGN heating, thereby reducing the cooling-to-heating ratio. In such a feedback loop, the key determinant of the cold-gas content is the efficiency of converting AGN power into thermal energy of the ICM. A higher conversion efficiency yields less cold gas for a given accretion rate, helping to suppress the cooling flow. We therefore expect `JetWind` to have the highest efficiency among the three models.

To test this, we computed the ratio of the time-integrated heating energy to the AGN energy released through jets and winds (excluding radiation). Radiative heating rate, given by $\propto n^2(T_C - T)F$, with $n$ the number density, $T$ the temperature of the gas, $T_C$ the Compton temperature of the AGN radiation, and $F$ the radiation flux, is most effective at small radii where both $n$ and $F$ are high but is insufficient to offset cooling at larger radii (*17,47*). Indeed, 85% of the radiative heating occurs within 1 kpc. By contrast, jet and wind energy is dissipated mainly through turbulence and shocks at larger scales.

The calculation of turbulent dissipation rate and shock heating rate are presented in Methods. Figure 5 shows the time evolution of turbulence dissipation rate and shock heating rate for the three models. The time-integrated turbulence dissipation energies for `JetWind`, `WindOnly` and `JetOnly` are $E_{\mathrm{tur,total}} = 4.6, 3.9, 6.0 \times 10^{61}$ erg, respectively. The converting efficiencies, defined as the fraction of the total released kinetic AGN energy that is dissipated by turbulent dissipation, are $29.6\%, 9.3\%$, and $5.2\%$, respectively. The respective shock-heating energies for the three models are $E_{\mathrm{shock,total}} \approx 5.1, 3.6, 5.7 \times 10^{61}$ erg, respectively, with the shock energy conversion efficiencies of $33.0\%$, $8.6\%$, and $4.9\%$. Combining the turbulent dissipation and shock heating yields total conversion efficiencies of $62.6\%, 17.9\%$, and $10.1\%$. Thus, `JetWind` indeed exhibits the highest efficiency, confirming our expectation and explaining why it alone eliminates the cooling flow.

A deeper question is why `JetWind` achieves such high efficiency. In our simulations, turbulence is mainly produced by the shear between jet, wind, and ambient ICM through the Kelvin-Helmholtz instability, cascading to smaller scales and converting kinetic to thermal energy. The turbulence amplitude in `JetWind` is the largest (see Methods), accounting for its superior energy-conversion efficiency.

## Discussion

Our simulations presented in this work are idealized and do not account for cosmological processes such as mergers and large-scale cosmological inflows. We note that the majority of clusters are actively accreting and have experienced recent merger events; therefore, caution is required when comparing our simulation results with observations. Nevertheless, some studies have suggested that mergers, cosmological inflows, and substructure evolution play only a minor role in suppressing cooling flows (*48,49*). In future work, we plan to systematically investigate these effects.

AGN feedback plays a fundamental role not only in alleviating the cooling flow problem but also in regulating galaxy formation and evolution more broadly. Jets constitute a primary channel for energy injection in AGN feedback, underscoring the importance of understanding how their energy is deposited into the host galaxy. Our results suggest that the coupling between jets and winds can provide an efficient mechanism for converting AGN kinetic power into effective feedback on the host system, particularly in massive halos where sustained AGN activity is present. This process therefore merits explicit consideration in future theoretical and numerical studies, depending on the physical regime and scientific goals being addressed.

## Materials and Methods

### Introduction to the MACER3D framework

The simulations were carried out using MACER3D, a comprehensive three-dimensional extension of the earlier two-dimensional MACER framework. The MACER project builds upon a series of foundational studies on the co-evolution of supermassive black holes and their host galaxies (*50,51*). The most recent 2D MACER framework (*33*) incorporates state-of-the-art prescriptions for AGN radiation and winds as functions of accretion rate, including radiative heating and cooling processes such as bremsstrahlung, Compton heating/cooling, photoionization heating, and line plus recombination continuum cooling.

Additional physical processes, such as star formation and stellar feedback, are also included. While the MACER framework technically supports the inclusion of cosmological inflow (*52*), this feature was explicitly disabled in this study. The inner boundary of the simulation domain is set smaller than the outer boundary of the black hole accretion flow. This setup allows the mass flux measured at the inner boundary to be reliably combined with black hole accretion theory to determine the precise mass accretion rate at the event horizon, which is the key parameter governing AGN power and feedback strength. MACER3D (*32*) further extends this framework by introducing physically motivated models for supernova feedback, radiative cooling, and metal enrichment. The implementation of the grid-based Athena++ code (*53*), utilizing a low-diffusion HLLC Riemann solver and a second-order Runge-Kutta (RK2) time integrator, enables robust three-dimensional modeling of non-axisymmetric instabilities and multiphase gas dynamics. Since no explicit viscosity is incorporated into the simulation setup, the resolved turbulence is effectively determined by the numerical diffusion of the solver (*53*), allowing for the treatment of complex physical features that were inaccessible to earlier two-dimensional axisymmetric versions (*33*). Radiative cooling is calculated via exact integration as described in (*32*).

Our simulations solve the three-dimensional hydrodynamic equations in spherical coordinates $(r, \theta, \phi)$ using the standard Euler form:

$$\frac{\partial \rho}{\partial t} + \nabla \cdot (\rho \mathbf{v}) = -\dot{\rho}_{\star}^{+}, \tag{1}$$

$$\frac{\partial \mathbf{m}}{\partial t} + \nabla \cdot (\mathbf{m}\mathbf{v}) = -\nabla p_{\text{gas}} + \rho \mathbf{g} - -\nabla p_{\text{rad}} - \dot{\mathbf{m}}_{\star}^{+}, \tag{2}$$

$$\frac{\partial E}{\partial t} + \nabla \cdot (E\mathbf{v}) = -p_{\text{gas}} \nabla \cdot \mathbf{v} + H - C - \dot{E}_{\star}^{+}, \tag{3}$$

where $\rho$, $\mathbf{m}$, and $E$ are the gas mass, momentum, and internal energy per unit volume, $\mathbf{v}$, $\mathbf{g}$ are the velocity and the gravitational field of the galaxy cluster, and $\dot{\rho}_{\star}^{+}$, $\dot{\mathbf{m}}_{\star}^{+}$ and $\dot{E}_{\star}^{+}$ denotes the sink terms of mass, momentum and energy due to star formation, respectively. The gas pressure is $p_{\text{gas}} = (\gamma - 1)E$ with adiabatic index $\gamma = 5/3$, and $\nabla p_{\text{rad}}$ represents the radiation force (*33*). Finally, $H$ and $C$ denote the radiative heating and cooling rates. The computational domain extends from an inner boundary of $r_{\text{in}} = 100$ pc to an outer boundary of $r_{\text{out}} = 2$ Mpc, a configuration designed to simultaneously resolve gas accretion flows near the SMBH and large-scale environmental effects within the galactic halo. We employ a fiducial resolution of $256 \times 64 \times 128$ cells. While the angular coordinates $(\theta, \phi)$ are uniformly discretized, the radial grid is logarithmically spaced with a constant scaling ratio $\Delta r_{i+1}/\Delta r_i = 1.038$, This grid configuration enhances the resolution in the cluster core, achieving a peak spatial resolution of ~ 5 pc at the inner boundary.

In the MACER framework, black hole accretion operates in two distinct modes, namely hot and cold mode, separated by 2% $L_{\text{Edd}}$ (*28*). In the cold mode, gas infalling through the inner boundary forms a thin disc at the circularization radius. The disc is continuously fed by the inflow, while mass is depleted through accretion onto the black hole and through winds launched from the disc surface. Solving the corresponding set of differential equations yields the black hole mass accretion rate $\dot{M}_{\text{BH}}$ (*33*). In this mode, the mass flux and velocity of the wind are determined from the observed statistical relation between AGN luminosity and outflow properties (*33*). The cold wind mass flux and velocity are

$$\dot{M}_{\text{W, C}} = 0.28 \left( \frac{L_{bol}}{10^{45}\ \text{erg s}^{-1}} \right)^{0.85} M_{\odot}\ \text{yr}^{-1}, \tag{4}$$

$$v_{\text{W, C}} = 2.5 \times 10^{4} \left( \frac{L_{bol}}{10^{45}\ \text{erg s}^{-1}} \right)^{0.4} \text{km s}^{-1}, \tag{5}$$

where $L_{\rm bol} = 0.1\,\dot{M}_{\rm BH}\,c^2$ is the bolometric luminosity. The mass flux of the wind adopts a bipolar angular distribution $\propto \theta$ with a half-opening angle $\theta_{\rm W,\,C} = 60°$ at the inner boundary.

Hot accretion flows have been extensively studied (*28*). In the hot mode, the accretion flow consists of an outer truncated thin disc and an inner hot accretion flow, with the truncation radius $r_{\rm tr} = 3r_s\left[\frac{2\times10^{-2}\dot{M}_{\rm Edd}}{\dot{M}(r_{\rm in})}\right]$ determined solely by the accretion rate of the thin disk. Magnetohydrodynamic simulations demonstrate that powerful winds are launched across the entire hot accretion region (*26-29*), a prediction increasingly supported by observations (*30,54-56*). Following the hot accretion flows theory (*29,33*), given the mass inflow rate $\dot{M}(r_{\rm in})$ at the inner boundary, the black hole accretion rate can be calculated by (*29*)

$$\dot{M}_{\rm BH} = \dot{M}(r_{\rm in})\,\left(\frac{2.5r_{\rm s}}{r_{\rm tr}}\right)^{0.42}. \tag{6}$$

Given the limited observational constraints of hot wind, MACER adopts wind properties derived from 3D GRMHD simulations (*29*). Specifically, the mass flux and velocity of hot wind are

$$\dot{M}_{\rm W,\,H} = \dot{M}(r_{\rm in})\,\left[1-\left(\frac{2.5r_{\rm s}}{r_{\rm tr}}\right)^{0.42}\right], \tag{7}$$

$$v_{\rm W,\,H} = 0.64\,v_{\rm k}(r_{\rm tr}), \tag{8}$$

where $\dot{M}(r_{\rm in})$ is the inflow mass flux at the inner boundary $r_{\rm in} = 100\,{\rm pc}$, and $r_{\rm s} = 2GM_{\rm BH}/c^2$ is the Schwarzschild radius. The wind velocity is determined by the Keplerian speed at the truncation radius $v_{\rm k}(r_{\rm tr})$ where most of the wind mass flux originates, beyond which the poloidal velocity of wind remains approximately constant (*33*). The angular distribution of the hot wind is restricted to 10°–50° and 130°–170° (*29*). The black hole accretion rate at the event horizon is then computed by combining the measured mass flux at the inner boundary of the simulation domain and truncation radius with the theory of wind from hot accretion flows (*33*).

The AGN releases radiation, winds, and jets. Previous MACER studies included only radiation and winds, while neglecting jets, which play a crucial role in addressing the cooling flow problem. In the next section, we describe in detail how jets are implemented in the present work. Once the properties of all AGN outputs are determined, they are injected at the inner boundary of the simulation domain, and their energy and momentum exchanges with the ICM are computed self-consistently. This approach avoids the use of parameterized, phenomenological feedback prescriptions commonly employed in earlier simulations.

## Modeling of AGN jet

In almost all existing sub-grid jet feedback models, the key jet parameters are treated as almost unconstrained free variables, whose values are not necessarily consistent with the constraints obtained from GRMHD simulations of black hole accretion and jet formation (*29,57-59*). The formation and properties of jets are largely determined by two factors: the black hole spin and the accretion mode (SANE or MAD). Current studies suggest that powerful jets are launched in the magnetically arrested disk (MAD) mode (*60,61*). We adopt a high dimensionless spin parameter, $a = 0.98$, consistent with spectral and timing constraints on massive, cluster-center AGN (*62,63*). Under these assumptions, the jet parameters are taken from the 3D GRMHD simulations of MAD accretion onto rapidly spinning black holes presented by (*29*). In that work, the jet (and wind) properties were derived using the "virtual test-particle trajectory" approach, which can more accurately separate turbulent motion from genuine outflow compared to the commonly used "streamline" method, thus providing more reliable jet (and wind) parameters. Since the outer boundary of the (*29*) simulations extends only to ~ $10^3 r_g$ (where $r_g = GM_{BH}/c^2 \sim 1.67 \times 10^{-5}$ pc is the gravitational radius), much smaller than the inner boundary of our simulation domain (100 pc ~ $6 \times 10^6 r_g$), extrapolation of the GRMHD results is required for implementation in MACER.

The first quantity to determine is the total jet energy flux. GRMHD numerical simulations and analytical studies show that, after launch, the Poynting flux of the jet is efficiently converted into kinetic and thermal energy within the acceleration zone (up to parsec scales) through magnetohydrodynamic processes and internal shocks (*57,64*). We assume that the total jet energy flux remains conserved during this process and adopt the value given by (*29*):

$$\dot{E}_{jet} = \dot{E}_{kin,jet} + \dot{E}_{th,jet} = 0.9\dot{M}_{BH}c^2, \quad (9)$$

where $\dot{E}_{kin,jet}$ and $\dot{E}_{th,jet}$ denote the kinetic and thermal components, respectively.

The second key parameter is the jet velocity. The mass flux-weighted jet velocity at $200\ r_g$ (~ $3.3 \times 10^{-3}$ pc) in obtained in (*29*) is $0.5\ c$. Since the inner boundary of our MACER simulations is much larger than the outer boundary of (*29*) and other typical GRMHD simulations of jet formations, we adopt the observed jet speed as the injection jet speed in our model. Although there is no direct measurement of the jet speed at ~100 pc, observational constraints are fairly tight. High-resolution millimetre VLBI observations of the jet in Perseus cluster reveal an apparent speed of $0.055 - 0.22\ c$ at ~1 pc scale (*65*). The jet viewing angle in this source is estimated to be $10° - 35°$. Furthermore, a detailed kinetic study of (*66*) indicates that the jet is subrelativistic and maintains an approximately constant speed from ~ 1 pc out to many parsecs, showing no sign of acceleration. The jet in the Perseus cluster is a relatively weak FR I jet that propagates through the compact core of the Perseus cluster. Combining the above information, we adopt the jet velocity at the inner boundary of our simulation domain to be:

$$v_{jet} = 0.1\ c. \quad (10)$$

Since we assume that the sum of kinetic power and the thermal power of the jet remains unchanged, we increase the thermal energy of the jet to satisfy this condition.

The third key parameter is the jet mass flux. To estimate this quantity, we have re-analyzed the original GRMHD simulation data presented in (*29*) and derived the mass flux of the jet at a distance of $1000\ r_g$ (~ $3.3 \times 10^{-3}$ pc):

$$\dot{M}_{jet} = 0.35\,\dot{M}_{BH}. \quad (11)$$

We assume that the mass flux remains roughly unchanged out to $6 \times 10^6\ r_g$ (100 pc) (*68*).

The fourth key parameter is the jet half-opening angle $\theta_j \equiv (R/Z)$, where $R$ and $Z$ are the half-radius and the distance from the black hole, respectively. There is no direct observational constraint on the value of this angle so we use theoretical result. Ref. (*29*) found that the jet radius follows a power-law relation with distance, $R = 1.01\ Z^{0.8}$. Using this profile, we calculate the half-opening angle of the jet at $\sim 10^4 r_g$ to be $\sim 7.5°$. Beyond this radius, up to $10^7 r_g$, observations indicate that the opening angle of the jet roughly remains constant (*67*). Therefore, we adopt a half-opening angle of the jet at the inner boundary of our simulation domain to be:

$$\theta_{\rm j} \approx 7.5^\circ. \tag{12}$$

This value is roughly consistent with the observational result reported by (*68*). The jet is initialized in the ghost zones at each time step using its prescribed mass flux and velocity, with its thermal power incorporated as the internal energy and is subsequently launched from the injection regions at the inner boundary.

Finally, although Ref. (*29*) found some angular dependence of velocity within the jet, Ref. (*67*) showed that this variation becomes increasingly flattened as the jet propagates outward. Accordingly, in the present work, we assume a uniform velocity across the jet cross section. No ad hoc jet precession is imposed.

## Comparison between MACER and other models

In this section we compare the MACER model with other AGN feedback models, including both idealized cooling flow simulations and large-scale cosmological simulations.

1) Accretion and feedback modes. According to the black hole accretion theory, accretion proceeds in two distinct modes, namely hot and cold modes, depending on the normalized black hole accretion rate. These two accretion modes naturally correspond to two feedback modes. Jet exists only in the hot mode while winds are expected to exist in both modes (*28*). The MACER model explicitly incorporates this two-mode accretion–feedback paradigm: in the hot mode, both jets and winds are self-consistently included, while in the cold mode only winds are present.
   Many existing idealized cooling flow simulations do not adopt this two-mode framework and instead implement only jet feedback (e.g., *10-19*).
   Cosmological simulations exhibit substantial diversity in their treatment of AGN feedback. EAGLE does not distinguish between hot and cold accretion modes and implements AGN feedback exclusively through isotropic thermal energy injection, without explicitly modeling jets or winds (*69*). Illustris-TNG adopts a two-mode feedback scheme but neither mode explicitly models jets or winds; instead, feedback is implemented phenomenologically (*70*). COLIBRE does not enforce a strict two-mode dichotomy. It includes both jet and wind feedback, but allows both components to operate simultaneously across different accretion states (*71*). SIMBA implements a two-mode framework, in which winds operate in the cold mode and jets are launched exclusively in the hot mode, but the two components do not coexist (*72*). Among current cosmological simulations, OBSIDIAN adopts a feedback structure most similar to MACER, allowing both jets and winds to coexist in the hot mode while including only winds in the cold mode (*73*).
2) Jet opening angle. In current cosmological simulations, AGN jets are generally implemented using sub-grid or phenomenological prescriptions, and their physical properties vary substantially among different models. In IllustrisTNG, collimated relativistic jets are not explicitly modeled; instead, the low-accretion-rate feedback mode is implemented as kinetic energy injection with weak or no explicit collimation, and without well-defined jet velocity, opening angle, or mass flux. SIMBA, COLIBRE, and OBSIDIAN include more directional outflows that are sometimes referred to as jets; however, their velocities, mass-loading factors, and opening angles are typically fixed or semi-empirical parameters chosen to reproduce galaxy-scale observables, rather than being derived from black hole accretion physics. For this reason, the following comparisons of jet properties focus primarily on idealized cooling flow simulations.
   In MACER, we adopt a jet half-opening angle of 7.5°, a value motivated by 3D

GRMHD simulations of black hole accretion and jet formation. The jet is assumed to be non-precessing. In other works, jets are usually modeled as bipolar precessing outflow, with assumed precessing angle $\theta_{\rm prec}$ ranging from 8.6° (*13,17*), to 15° (*14*), 25° (*10,19*), and 30° (*18*), typically with an assumed precession period of ∼10 Myr. In these models, the outflow is injected along the z-axis from two parallel "jet launching planes" located at a height $h_j$ from the black hole, each with a radius $R_j$ . This setup corresponds to an effective jet half-opening angle of $arctan\ (R_{\rm j}/h_{\rm j}) \approx \theta_{\rm j} = 30°$ (*12*), 37° (*13*), 45°(*17*), 51°(*14*), and 66° (*18*). These values are substantially larger than those inferred from observations or GRMHD numerical simulations of jet formation. Moreover, the injected jet velocity in these models is purely along the z-direction, unlike the radially oriented velocity field expected for physically launched jets. Although some of these studies adopt relatively large jet opening angles that may partially resemble a combined wind–jet outflow, other jet properties, such as velocity, density, and their angular dependence, remain fundamentally different from the physically motivated wind+jet properties implemented in MACER.

3) Jet mass flux. In MACER, the jet mass flux is prescribed as a function of the black hole accretion rate, based on the 3D GRMHD simulations of black hole accretion. In contrast, many other works treat the jet mass flux as a free parameter, often without a clear physical justification for the adopted value. As a result, the assumed jet mass-loading factors vary widely, ranging from $\dot{M}_{\rm jet}/\dot{M}_{\rm BH} \sim 1$ (*13,14,17,18*), to $\sim 0.432$ (*10,19*), and down to $\sim 0.006$ (*12*).
4) Jet velocity. In our work, the jet velocity (0.1 $c$) is constrained by a combination of radio observations of the jet in the Perseus cluster and results from 3D GRMHD simulations of black hole accretion, as we have explained in the last section. In previous studies (10-14,17-19), the adopted jet velocities span a wide range, from 0.16 $c$ (*10,19*) and 0.1 $c$ (*12,13*), down to 0.045 $c$ (*14,17*) and 0.033 $c$ (*18*). Our adopted jet velocity therefore lies well within the range assumed in previous works.
5) Wind launched from the black hole accretion flow. In MACER, the properties of wind, including the opening angle, velocity, density, and their angular and accretion-rate dependences are largely taken from GRMHD simulations of black hole accretion in the hot mode and observations in the cold mode. These properties differ substantially from those of the jet. In particular, the angular distribution of outflow velocity plays a crucial role. As demonstrated in the present work, a physically consistent combination of wind and jet velocities naturally generates strong turbulence, whose dissipation efficiently converts AGN kinetic power into thermal energy of the ICM.
As noted above, most idealized cooling flow simulations do not include winds. Among cosmological simulations, EAGLE does not explicitly model winds or jets and it implements AGN feedback exclusively through isotropic thermal energy injection (*69*). IllustrisTNG also does not explicitly model winds or jets. The hot mode employs kinetic energy injection, but this kinetic feedback is implemented phenomenologically, without explicitly modeling collimated jets or physically motivated AGN winds launched from the accretion flow. In the cold mode, there is no wind either and the AGN feedback is implemented in a way similar to EAGLE (*70*). Although COLIBRE includes a component referred to as "wind", its implementation is fundamentally different from the physically motivated accretion-driven winds in MACER. In COLIBRE, the wind is implemented phenomenologically as part of the AGN kinetic feedback, whereas in MACER the wind properties are directly constrained by black hole accretion theory and GRMHD simulations (*71*). SIMBA launches winds in the cold mode, but the wind is implemented phenomenologically, with its properties calibrated to observations rather than derived from black hole accretion physics (*72*).

Although OBSIDIAN is the most physically comparable to MACER, the wind and jet properties are not directly constrained by GRMHD simulations but are implemented using phenomenological sub-grid prescriptions tied to the accretion rate and feedback efficiency (*73*).

6) Black hole accretion rate. In MACER, the value of the inner boundary of the simulation domain is smaller than the outer boundary of the accretion flow (see Fig. 2). This enables a physically motivated calculation of the accretion rate at the black hole horizon by combining the mass flux at the inner boundary with black hole accretion theory (See *33* for details).
In contrast, in most other works, the accretion rate is only crudely estimated. In idealized cooling flow simulations it is often calculated by dividing the total amount of cold gas within an arbitrarily assumed radius by an assumed accretion timescale, with little justification for the adopted radius (*12-14,17-19*). In cosmological simulations, Bondi or modified Bondi accretion rate are commonly applied in both hot and cold accretion modes to estimate the black hole accretion rate (*69-71,73*) or hot mode for *SIMBA* (*72*). For the cold mode accretion, SIMBA models the gas inflow rate driven by disc gravitational instabilities while OBSIDIAN considered cold gas accretion using a way similar to idealized cooling flow simulations. We emphasize that if the Bondi radius is unresolved in the simulation, it is not correct to apply the Bondi or modified Bondi formula, otherwise the errors can be as large as 100 or even much higher (e.g., *75,76*).
7) Interaction between AGN outflow and ICM/ISM. In MACER, both wind and jet are injected at the inner boundary, and their momentum and energy interaction with the ICM/ISM is then self-consistently calculated. In contrast, in the "thermal feedback mode'' adopted in many cosmological simulations, AGN feedback is typically implemented in a phenomenological manner, namely as isotropic heating of gas within a certain distance from the black hole (*69-71*). In such implementations, both the fraction of AGN energy deposited into the gas and the deposition radius are free parameters with limited physical motivation. More importantly, the momentum feedback effect associated with AGN winds, which may play a more important role than pure energy feedback (*76*), is commonly neglected. Furthermore, to mitigate numerical overcooling, feedback energy is commonly stored in a reservoir until a sufficient amount is accumulated, an approach that again lacks a clear physical basis.
8) Resolution. The highest resolution of MACER is achieved at the small radii, which is $\sim 5$ pc. The highest resolution of other idealized isolated cluster simulations are $\sim 15-20$ pc (*12*), $15-60$ pc (*13*), $30-160$ pc (*16*), several hundred parsecs (*17-20*) and 1.95 kpc (*14*). In cosmological simulations, spatial resolution is typically constrained by gravitational softening lengths ranging from 0.3 kpc in state-of-the-art runs like TNG50 (*31*) to $\sim 0.7$ kpc in EAGLE (*69*) and SIMBA (*72*).

## Initial conditions of the simulations

We adopt the archetypal, relaxed cool-core galaxy cluster Perseus as a reference system for our simulations, while noting that our models are not intended to reproduce this specific cluster in detail due to the lack of cosmological effects such as merger and cosmological inflow in our simulations. Following the work of (*13*), we initialize the ICM as a hydrostatic sphere of gas within a static spherical gravitational potential. The gravitational potential comprises three components: an NFW halo, the stellar mass profile of the BCG and the SMBH. The NFW halo density profile is (*77*)

$$\rho(r) = \frac{\rho_0}{\frac{r}{R_S}\left(1+\frac{r}{R_S}\right)^2}, \tag{13}$$

where $\rho_0$ is found to be $8.42 \times 10^{-26}\ g\ cm^{-3}$, and the scale radius $R_s = 351.7$ kpc. The brightest cluster galaxy (BCG) is treated as a fixed potential (*36*) with the stellar acceleration:

$$g_*(r) = \left[\left(\frac{r^{0.5975}}{3.206\times10^{-7}}\right)^{0.9} + \left(\frac{r^{1.849}}{1.861\times10^{-6}}\right)^{0.9}\right]^{-1/0.9}. \quad (14)$$

The SMBH in the center of the cluster is treated as a point mass of $M_{\rm SMBH} = 3.4 \times 10^8\ M_\odot$ (*37*). Combining the black hole mass and the typical temperature of the hot ICM gas, the inner boundary $r_{in}$ of our simulations is set to $100\ pc$, which is roughly equal to the outer boundary of the accretion flow.

We assume an ideal gas law for the ICM with $\gamma = 5/3$, and that the ICM is initially in hydrostatic equilibrium with the gravitational potential which includes the contribution from the NFW halo, the BCG and the supermassive black hole. We do not include any initial rotation or perturbation in the gas. The hydrostatic gas in the halo is initialized with an electron density profile of the form (*41*)

$$n_e(r) = \left(\frac{0.0192}{1+\left(\frac{r}{18}\right)^3} + \frac{0.046}{\left[1+\left(\frac{r}{57}\right)^2\right]^{1.8}} + \frac{0.0048}{\left[1+\left(\frac{r}{200}\right)^2\right]^{1.1}}\right) \mathrm{cm}^{-3}, \quad (15)$$

where $r$ is the distance to the cluster center in kpc.

Following (*13*), the initial ICM temperature profile within the central ($r < 300$ kpc) is constrained by observations (*78*) as

$$T = 7\,\frac{1+(r/71)^3}{2.3+(r/71)^3}\ \mathrm{keV}, \quad (16)$$

while at larger radii ($r > 300$ kpc) we adopt the universal profile proposed by (*79*):

$$T = 9.18\ \left(1 + \frac{3\,r}{4880}\right)^{-1.6}\ \mathrm{keV}. \quad (17)$$

Combining with the ICM electron density profile, we have $r_{200} = 1.83$ Mpc and $M_{200} = 7.47 \times 10^{14} M_\odot$ defined as the radius within which the mean enclosed density is 200 times the critical density. Based on this, the outer boundary of our isolated simulations is set to approximately one virial radius as 2 Mpc, which is slightly larger than $r_{200}$ to mitigate the effects of boundary conditions. The power-law index of the last term is slightly steepened so that the density profile at large radii is more consistent with cosmological simulations as well as the observations of the outskirts of Perseus (*37*). Since we focus on the cluster core within the cooling radius, beyond which both cooling and dynamical timescales are very long, the exact value of the index is not very important. We do not include initial rotation in the gas.

## Three simulation models

In contrast to previous cluster scale simulations that predominantly investigated single-channel feedback mechanisms, MACER3D introduces a framework incorporating a multi-channel feedback scheme, including AGN jets as well as hot and cold winds. To systematically assess the importance of properly configuring multi-channel feedback in suppressing ICM cooling, we set up three simulations. The `JetWind` simulation includes both AGN jet and hot wind. For comparison, the `JetOnly` and `WindOnly` simulations disable the hot wind ($\dot{M}_{\rm W,\,H} = 0$) and the jet ($\dot{M}_{\rm Jet} = 0$), respectively, while keeping other feedback components (e.g., cold wind) active. All three simulations retain the cold wind at high accretion rates. Importantly, when comparing `JetOnly` and `WindOnly` with `JetWind`, the total feedback power under the same black hole mass accretion rate remains nearly identical; for example, in the `WindOnly` simulation, we set $\dot{M}_{\rm Jet} = 0$, but redistribute the jet power into the hot wind channel, and vice versa. This ensures that removing a specific feedback channel does not reduce the overall AGN output power, thereby allowing a fair comparison of the feedback effects for three simulation sets.

## Acknowledgments

We thank the two anonymous referees for their constructive comments and suggestions, which significantly improved the manuscript. Numerical calculations were run on the CFFF platform of Fudan University, the supercomputing system in the Supercomputing Center of Wuhan University, and the High Performance Computing Resource in the Core Facility for Advanced Research Computing at Shanghai Astronomical Observatory.

**Funding:**
National Natural Science Foundation of China grant 12133008 (FY)

National Natural Science Foundation of China grant 12192220 (FY)
National Natural Science Foundation of China grant 12192223 (FY)
National Natural Science Foundation of China grant 12361161601 (FY)
National Natural Science Foundation of China grant 12522301 (FY)
National Natural Science Foundation of China grant 12233005 (HX)
National Key R&D Program of China grant 2023YFB3002502 (SJ)
China Manned Space Project grant CMS-CSST-2025-A08 (FY)
China Manned Space Project grant CMS-CSST-2025-A10 (FY)
National SKA Program of China grant 2025SKA0130100 (FY)

**Author contributions:**

Conceptualization: FY
Methodology: FY, SJ, AH, MG
Software: AH, FY, SJ, MG, HJX
Validation: AH, FY, SJ, MG, YL, HGX, MS, HJX, YZ
Formal analysis: AH, FY, SJ, MG, YL, HGX
Investigation: AH, MG, FY, SJ
Resources: FY, SJ
Data curation: AH, MH
Writing—original draft: AH, FY
Writing—review & editing: AH, FY, SJ, MG, YL, HGX, MS, HJX, YZ
Visualization: AH, FY, SJ
Supervision: FY, SJ
Funding acquisition: FY, SJ, HGX
Project administration: FY

**Competing interests:** Authors declare that they have no competing interests.

**Data and materials availability:** All analysis scripts and the processed data underlying the figures are archived on Dryad (https://doi.org/10.5061/dryad.9ghx3ffz7). Primary 3D simulation datasets are available on Zenodo (https://doi.org/10.5281/zenodo.18805564).

## Figures and Tables

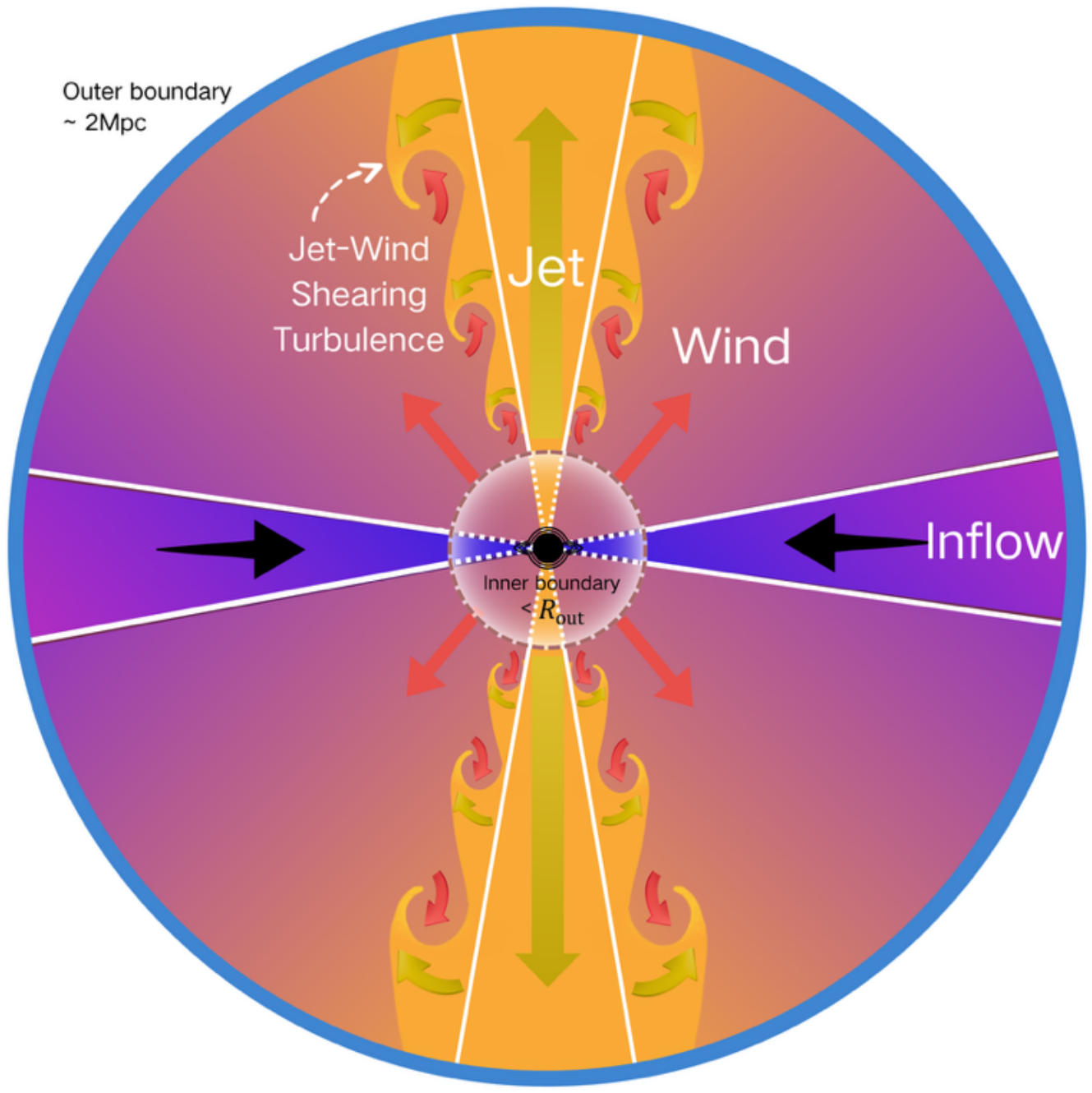

**Fig. 1. Schematic of the fiducial `JetWind` model.** The dashed circle marks the simulation's inner boundary, located within the outer boundary of the accretion flow $R_{\mathrm{out}}$, enabling self-consistent calculation of the black hole accretion rate and accurate coupling of AGN outputs to the intracluster medium (ICM). Jet and wind parameters are taken from GRMHD simulations. Shear between them triggers Kelvin–Helmholtz instabilities that drive turbulence, whose dissipation converts jet kinetic energy into ICM heat, suppressing the cooling flow.

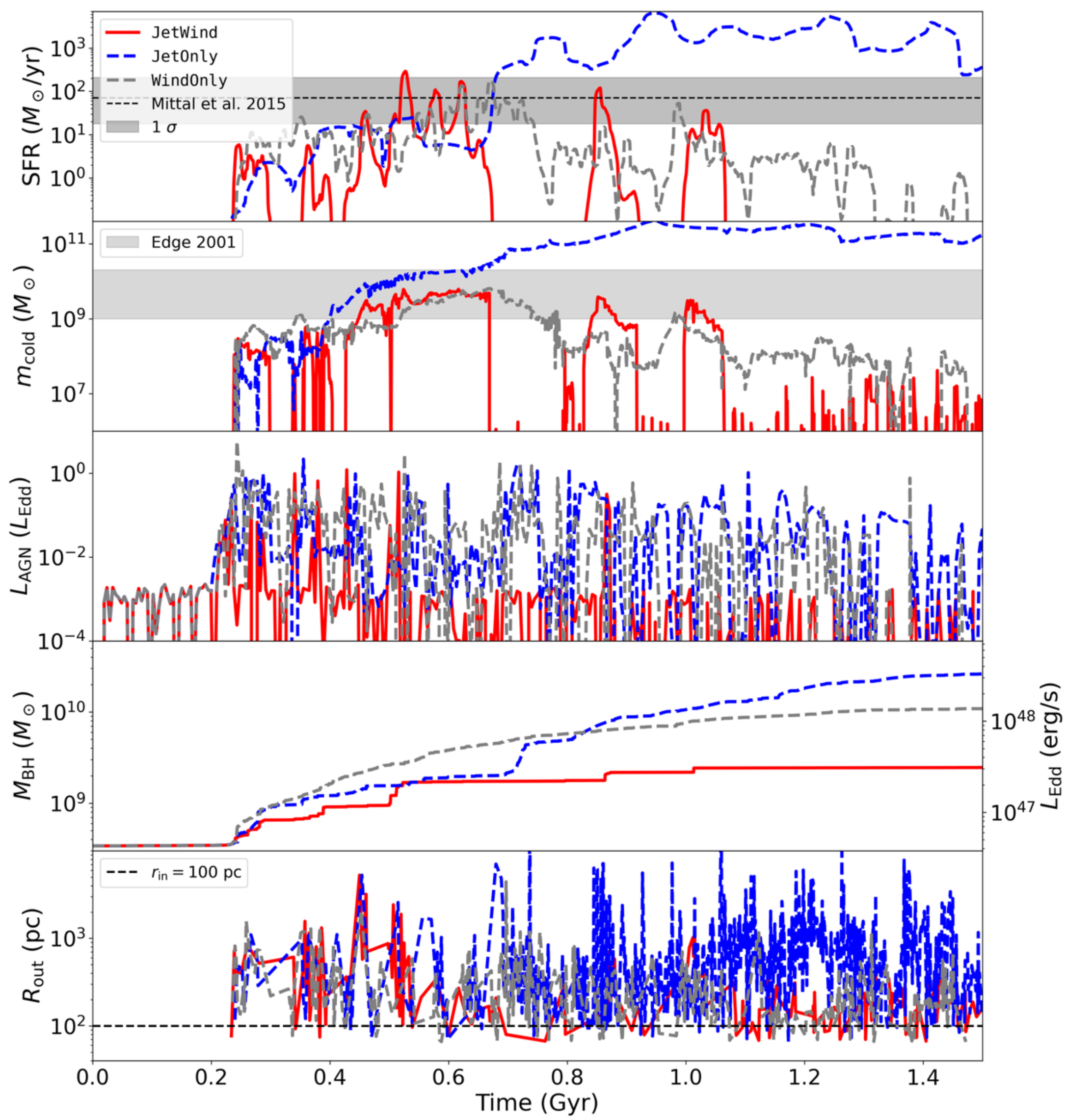

**Fig. 2. Time evolution of key quantities in the three models.** Results are shown for the `JetWind` (solid red), `JetOnly` (dashed blue), and `WindOnly` (dashed grey) models. *Top:* star formation rate (SFR); the black dashed line and grey shaded region mark the SFR of NGC 1275 (the Perseus BCG) and its 68 % confidence interval (*34*). *Second:* cold gas mass; the grey band denotes the molecular gas mass observed in Perseus (*35*). The observational values for SFR and cold gas mass represent upper limits for Perseus-like clusters. *Third:* AGN bolometric luminosity normalized by the Eddington luminosity. *Fourth:* black hole mass and Eddington luminosities; *Bottom:* evolution of the mass flux-weighted outer boundary of the accretion flow $R_{\rm out}$.

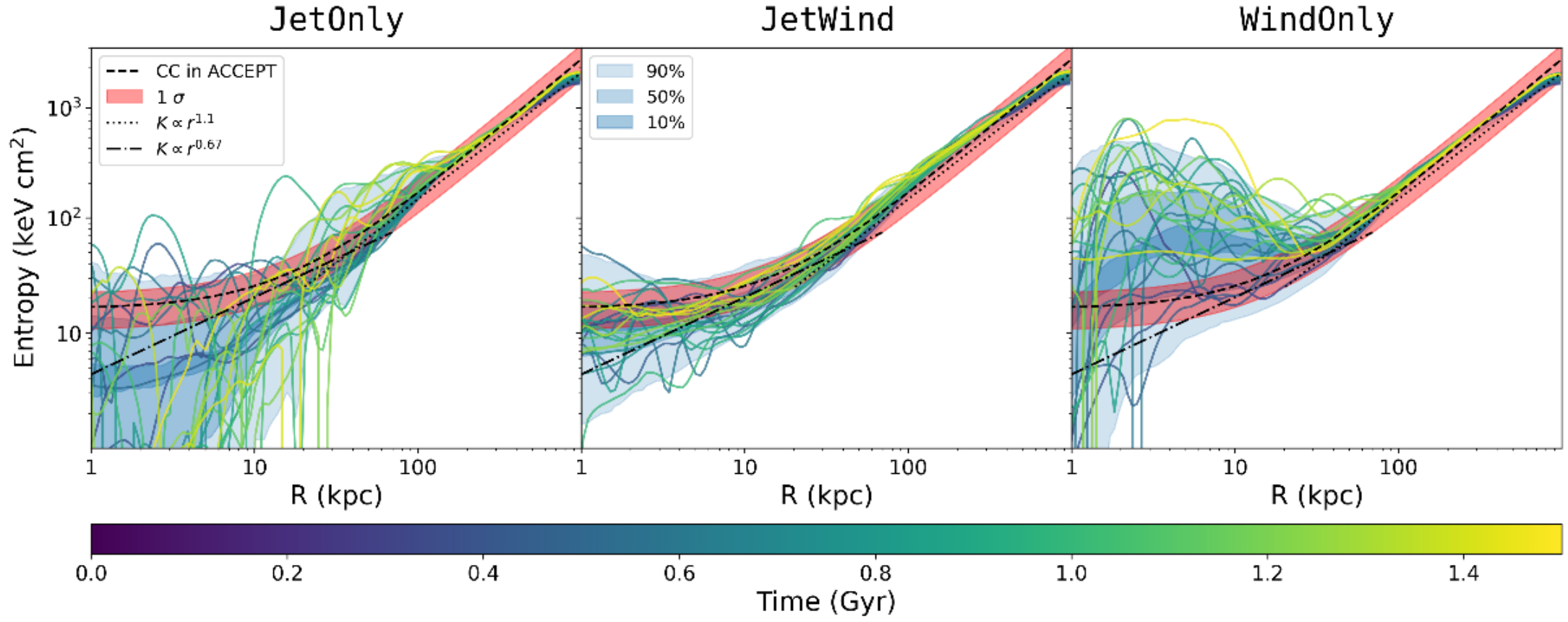

**Fig. 3. Radial profiles of entropy of the intracluster medium, color-coded by time**. The shaded regions denote the 10th–90th percentile distribution of the simulated profiles. Dashed black lines and red bands indicate the median and 1$\sigma$ observed profiles from the ACCEPT cluster sample (*39*). The dot-dashed and dotted lines represent the $K \propto r^{0.67}$ and $K \propto r^{1.1}$ power-law fits (*40*). Notably, only the `JetWind` model reproduces observed entropy profiles throughout most of its evolution.

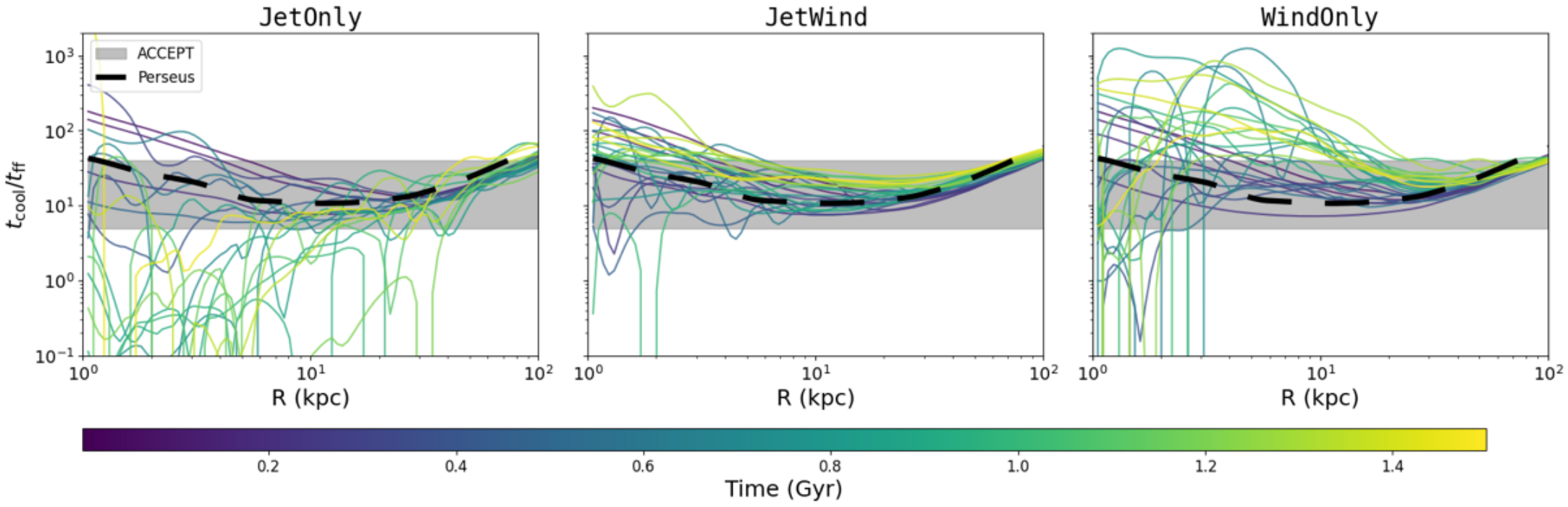

**Fig. 4. Radial profiles of the density-weighted cooling-to-free-fall time ratio.** Colored curves indicate simulation times, as shown by the color bar. The gray shaded regions mark observational ranges for cool-core clusters from the ACCEPT database (*44*), while the black dashed line shows Perseus data (*35,78*). Only the `JetWind` model maintains $t_{cool}/t_{ff}$ values consistent with observations across the core region.

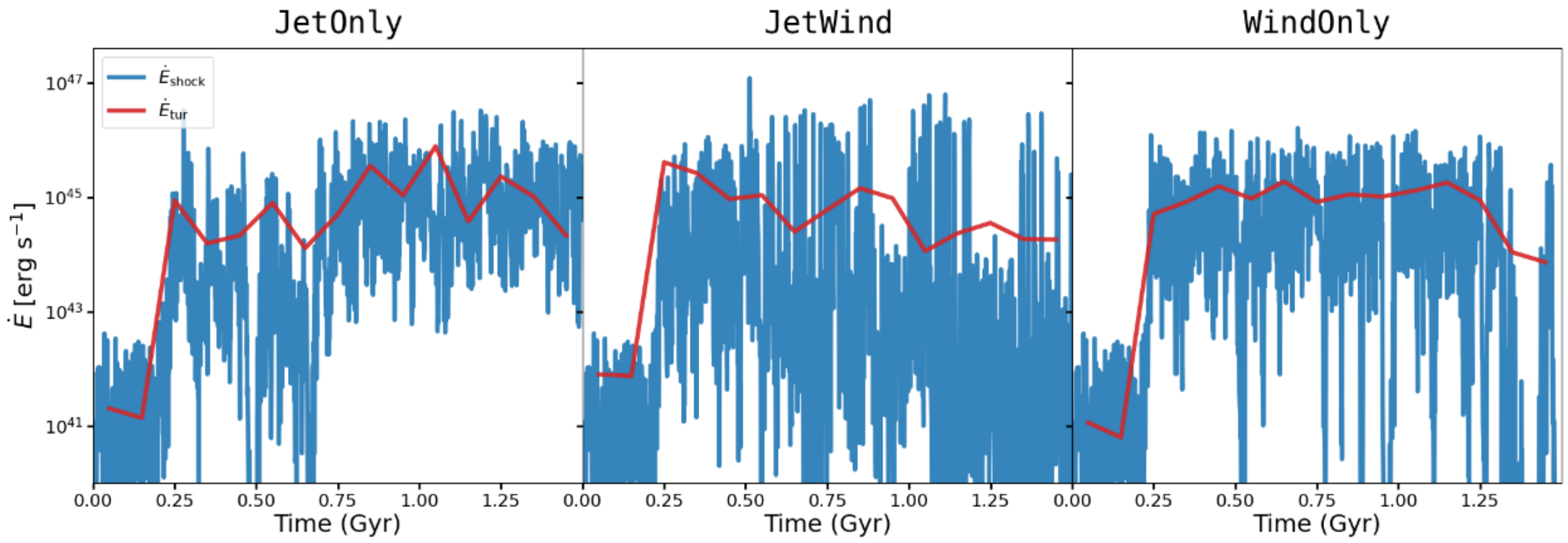

**Fig. 5. Time evolution of turbulence dissipation rate and shock heating rate.**

**Supplementary Materials**

Turbulent dissipation rate

Our approach to quantifying this process follows the methodology of (*18*). To avoid the contamination from the radial bulk motions directly injected by the AGN, we compute the velocity power spectrum using the velocity dispersion of $v_\theta$. This choice minimizes the impact of large-scale radial flows and ensures that the derived spectrum more faithfully represents the turbulent velocity field. We define the velocity fluctuation as

$$\delta v(r,\theta,\varphi) = v_\theta(r,\theta,\varphi) - \overline{v_\theta}(r)\,, \tag{S18}$$

where $\overline{v_\theta}(r)$ is the mean velocity at each radius. To avoid the influence of bulk motions in the $\theta$ direction caused by the lateral adiabatic expansion of hot bubbles generated by the outflows, the mean velocity and the velocity dispersion are computed separately in the regions with $\theta > \pi/2$ and $\theta < \pi/2$. This separation effectively reduces the contamination from large-scale coherent flows associated with bubble expansion and allows the resulting velocity statistics to better capture the underlying turbulent motions. We excluded the r-direction velocity dispersion because gravity wave damping in stratified media (like the ICM) substantially suppresses radial turbulence at certain positions, making theta-direction dispersion dominant. We also omitted phi-direction dispersion since our simulation is rotationally symmetric without artificial angular momentum.

Over the entire simulation time span of 1.5 Gyr, we obtain the time-averaged values of the velocity amplitude. Next, we perform a Fast Fourier Transform (FFT) of the one-component velocity amplitude to obtain the power spectral density $E_k(k)|_r$ at each fixed radius $r$, from which the velocity fluctuation field in frequency space can be derived as

$$\delta v(k)|_r = \sqrt{E_k(k)|_r k}, \tag{S19}$$

where for a given radius $r$, the wavenumber range is $k|_r \in \left(\frac{1}{r\mathrm{d}\theta}, \frac{1}{\pi r}\right)$ with $\mathrm{d}\theta$ denoting the minimum angular resolution in the $\theta$ direction. Since the wavenumber spaces corresponding to adjacent radii partially overlap, the resulting complete power spectrum should represent the dominant frequency in each mode. Therefore, we adopt a logarithmic binning scheme over the entire wavenumber range, selecting the maximum power spectral density within each $k$-bin, from which the corresponding velocity dispersion is then computed. Finally, by combining the velocity power spectra from all radial shells, we construct the complete power spectrum over the full wavenumber range $\delta v(k)$.

Similar to (*18*), we also assume that the turbulent energy density, $\frac{1}{2}\bar{\rho}\delta v(k_{\mathrm{inertial}})^2$, associated with turbulent eddies of size $\frac{1}{2}\bar{\rho}\delta v(k_{\mathrm{inertial}})^3 L \sim 1/k_{\mathrm{inertial}}$ , cascades down to smaller eddies on approximately one eddy turnover time, $t_{\mathrm{turn}} = (k\delta v)^{-1}$. Finally, the expression for the turbulent dissipation rate density is given by

$$\dot{e}_{\mathrm{tur}} = \frac{1}{2}\bar{\rho}\delta v(k_{\mathrm{inertial}})^3 k_{\mathrm{inertial}} \tag{S20}$$

where $\bar{\rho}$ denotes the mean density within the spatial region (30 kpc) used to compute the velocity power spectrum, and $k_{\mathrm{inertial}}$ is the wavenumber within the inertial range of a fully developed Kolmogorov turbulence power spectrum. The total turbulent dissipation power is then evaluated as $\dot{E}_{\mathrm{tur}} = \dot{e}_{\mathrm{tur}}\,V$, where $V$ represents the volume of the spherical region with a radius of 30 kpc.

Theoretically, the turbulent dissipation rate within the inertial range is independent of $k_{\mathrm{inertial}}$. To better estimate energy dissipation through energy cascading across different scales, the final turbulent dissipation power in each of the three simulations (`JetWind`, `JetOnly` and `WindOnly`) is obtained by averaging the dissipation rates computed over all wavenumbers within their respective inertial ranges of the power spectrum. This averaging procedure reduces the uncertainty introduced by numerical noise or fluctuations at individual wavenumber modes. This method provides a direct link between the simulated turbulence, energy dissipation, and the resulting heating rate in the ICM.

Shock heating rate

To calculate the shock heating, we adopt an approximate estimator of the shock heating rate, which is applied in post-processing. Our approach follows methods developed in previous works (*14,18*), where shocks are identified through entropy, pressure, and density jumps across consecutive snapshots. Once shocked cells are detected, the associated heating rate is estimated based on the entropy increase, providing a lower-limit measure of the energy dissipated by weak shocks into the thermal component of the ICM.

The pressure jump across a shock can be expressed as

$$\frac{\Delta P}{P} \equiv \frac{P_2 - P_1}{P_1} = \frac{2\gamma}{\gamma+1}\, y, \tag{S21}$$

where $\gamma = 5/3$ is the adiabatic index, and $P_1$ and $P_2$ are the pre-shock and post-shock pressures, respectively. The dimensionless parameter $y = M^2 - 1$ is related to the shock Mach number $M$, given by

$$M = \sqrt{\frac{\rho_1 v_s^2}{\gamma P_1}}, \tag{S22}$$

with $\rho_1$ the pre-shock density and $v_s$ the shock velocity. According to the classical Rankine-Hugoniot (RH) relations, the corresponding density jump across the shock is

$$\delta_\rho \equiv \frac{\rho_2}{\rho_1} = \frac{(\gamma+1)(y+1)}{2+(\gamma-1)(y+1)}. \tag{S23}$$

In the weak shock limit, the entropy jump across a shock can be approximated as

$$\mathrm{d}s \approx \frac{2\gamma k_B}{3(\gamma+1)^2 \mu m_H}\, y^3, \tag{S24}$$

where $\mu$ is the mean molecular weight, $m_H$ is the mass of a hydrogen atom, and $k_B$ is the Boltzmann constant.

In practice, we compute the pressure and density differences between consecutive snapshots separated by $\Delta t = 1$ Myr at the same spatial grid cells. Cells are flagged as shock-heated if they simultaneously satisfy the following conditions: (i). $0.2 < \Delta P/P < 40$, which corresponds to weak shocks in the Mach number range $1.08 < M < 5.7$; (ii). $ds > 0$, ensuring the presence of local heating associated with an entropy increase; (iii). $\rho_2/\rho_1 \geq \delta_\rho$, requiring the density jump to be consistent with the RH relations, thus confirming that the flagged cells trace genuine shocked regions. For the cells flagged as shock-heated, the shock heating rate density is estimated as

$$\dot{e}_{\mathrm{shock}} = \frac{\rho_1 T_1\, ds}{\Delta t}. \tag{S25}$$

The total shock heating rate is then obtained by summing over the shocked volume:

$$\dot{E}_{\mathrm{shock}} = \dot{e}_{\mathrm{shock}} V, \tag{S26}$$

with $V$ denoting the corresponding cell's volume.

Different from (*18*), our shock heating estimator adopts a higher time resolution of 1 Myr, which enhances our ability to capture the contribution of strong shocks. Moreover, in condition (i) we additionally choose 40 as the upper limit for the pressure jump to ensure the validity of the weak-shock approximation used in the entropy jump equation, and to exclude the contribution of strong shocks ($\Delta P/P \gg \frac{\gamma+1}{\gamma-1}$). In condition (ii), we further impose a more stringent density jump constraint to identify shocked regions, which ensures that our estimate does not overestimate the level of shock dissipation. This procedure provides an approximate yet practical measure of the energy dissipation associated with weak shocks driven by AGN jets and winds in the ICM.

KH instability driven by the jet-wind shear in the JetWind model produces the strongest turbulence among the three models

The shear among jet, wind, and ICM results in Kelvin-Helmholtz (KH) instability, which can potentially grow and cause turbulence. In the following, we compare the strength of turbulence produced in the three models.

Figure S1 shows the radial velocity ($v_r$) profiles as a function of polar angle ($\theta$) time-averaged over the hot mode phase for the three models. In each plot, different lines represent the radial velocity $v_r$ at various radii ranging from 1 kpc to 30 kpc. The `JetWind` model exhibits the largest velocity gradients, indicating persistent strong shear flows throughout the entire angular domain. In contrast, the `WindOnly` simulation displays more moderate shear profiles and the `JetOnly` simulation demonstrates even flatter velocity profiles, suggesting that isolated jet feedback is considerably less effective in generating the shear flows necessary for turbulence development. For the `JetOnly` case, the cold wind launched prior to the transition to the hot accretion mode produces strong outward radial velocities in the equatorial region. As a result, a significant outflow radial velocity persists even after time-averaging over the hot-mode phase.

Based on these velocity distributions, we can estimate the growth rate of the KH instability using the relation $\gamma \sim \chi^{1/2} k v_{\rm shear}$ , where $\chi$ denotes the density contrast between adjacent shear layers, $k$ denotes the perturbation wavenumber, and $v_{\rm shear}$ corresponds to the velocity differential across the shear interface. While the density contrast $\chi$ across the angular direction is comparable among three models, the shear velocity $v_{\rm shear}$ differs substantially. In `JetWind`, $v_{\rm shear} \sim v_r$, substantially exceeding that of the `JetOnly` simulation where $v_{\rm shear} \ll v_r$. This difference results in different timescale relationships: $t_{\rm KH} \sim t_{\rm dynamic}$ in `JetWind` versus $t_{\rm KH} \gg t_{\rm dynamic}$ in `JetOnly`, where $t_{\rm dynamic} \sim r/v_r$ represents the characteristic timescale for the propagation of the outflow. Consequently, the shear flows in the `JetWind` model generate much stronger turbulence, whose dissipation leads to more efficient heating.

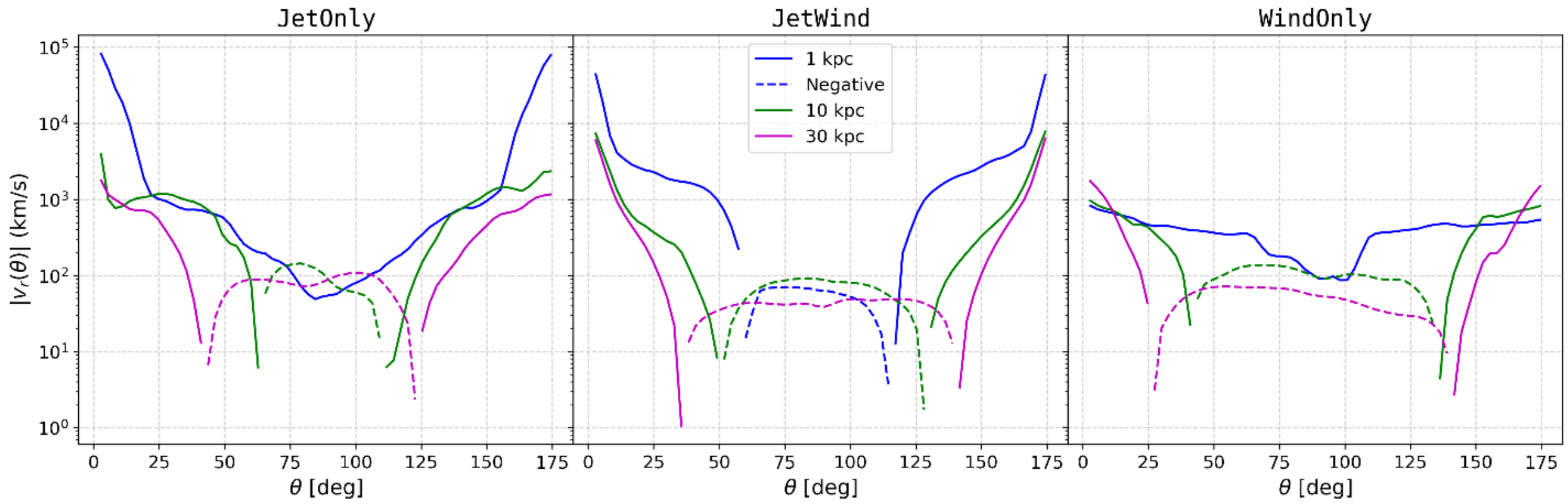

**Fig. S1.**

Radial velocity $v_r$ as a function of polar angle $\theta$, time-averaged over the hot mode phase for the three models. Different colors denote velocity profiles at various radii. The `JetWind` model exhibits the strongest velocity shear, which in turn drives the most vigorous KH instabilities and turbulence.

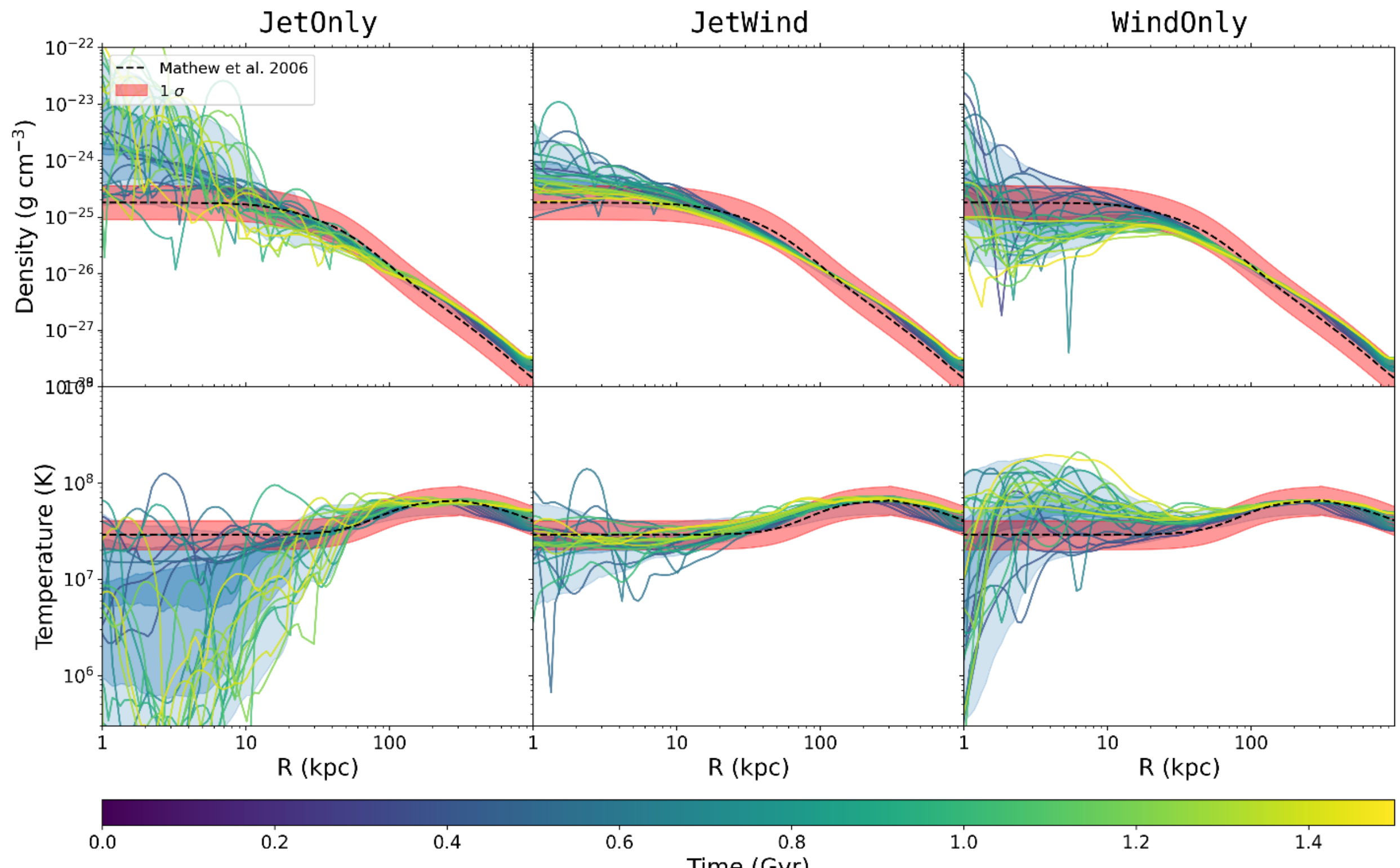

**Fig. S2.**

Radial profiles of the intracluster medium. Density (top) and temperature (bottom) profiles are shown for the three models, color-coded by time. Shaded areas mark the 10th–90th percentiles. The black dashed line and red shaded region shows Perseus data (*41,78*). Again, only the `JetWind` model's density and temperature profile primarily meets the observations.